\documentclass[12pt]{iopart}

\usepackage{lineno}
\usepackage{IEEEtrantools}

\expandafter\let\csname equation*\endcsname\relax

\expandafter\let\csname endequation*\endcsname\relax

\usepackage{amsmath}

\usepackage[dvipsnames]{xcolor}
\usepackage{graphicx}

\newcommand{\G}[1]{G_{#1}}
\newcommand{\g}[1]{g_{#1}}
\newcommand{\h}{\bar{H}}
\newcommand{\x}{\tilde{x}}

\begin{document}


\title{Instability of the cosmological DBI-Galileon in the non-relativistic limit}

\author{C. Leloup$^{1,2}$, L. Heitz$^{3}$ and J. Neveu$^{3,4}$}

\address{$^{1}$ Université Paris-Cité, CNRS, Astroparticule et Cosmologie, 75013 Paris, France}
\address{$^{2}$ Kavli Institute for the Physics and Mathematics of the Universe (Kavli IPMU, WPI), UTIAS, The University of Tokyo, Kashiwa, Chiba 277-8583, Japan}
\address{$^{3}$ Université Paris-Saclay, CNRS, IJCLab, 91405, Orsay, France}
\address{$^{4}$ Sorbonne Université, CNRS, Université de Paris, LPNHE, 75252 Paris Cedex 05, France}

\vspace{10pt}

\begin{center}
\begin{abstract}

The DBI-Galileon model is a scalar-tensor theory of gravity which is defined as the most general theory of the dynamics of a 4D brane embedded in a 5D bulk. It is of particular interest as it introduces only a few free parameters, all with a clear physical meaning, such as the brane tension which is related to the cosmological constant. From the tight constraints on the gravitational waves speed, we are naturally led to consider the non-relativistic limit of the model where the kinetic energy of the brane is small compared to its tension, that we study in the context of late-time cosmology. We show that the DBI-Galileon in the non-relativistic limit is an expansion around General Relativity (GR) which can be expressed as a shift-symmetric Horndeski theory. We develop the description of this theory at the background and perturbation level. However, by studying the scalar and tensor perturbations around a flat FLRW background, we find that they contain a ghost degree of freedom leading to a fatal instability of the vacuum for every combination of the free parameters. In order to avoid it in the more general cases of shift-symmetric Horndeski theories, we emphasize which of the Horndeski terms are in competition to produce this instability.
\end{abstract}
\end{center}

\setlength{\mathindent}{0pt}

%
%
%
%
%

\section{Introduction}

Dark energy has been modelled by a large variety of theories. To model a late-time acceleration of the Universe expansion, scalar-tensor theories rely on the introduction of additional scalar fields whose dynamics are determined by arbitrary parametric functions, potentials and/or coupling (see \textit{e.g.} \cite{Brax_2018}). 
In particular, the class of Horndeski theories 
contains all modified gravity models with a single additional scalar field leading to second-order equations of motion  \cite{Horndeski:1974wa,Deffayet:2011gz}. Through four arbitrary functions of the scalar field and its kinetic energy, they lead to a huge variety of models and phenomenological behaviours. Extensions of Horndeski theories to scalar-tensor theories 
with equations of motion of higher orders have also been explored \cite{Langlois:2015skt,Langlois:2018dxi}. 

Among this wide class of models, some arise from first physical principles or symmetry arguments. For instance, the Galileon model \cite{Nicolis:2008in} and its covariant extension \cite{Deffayet:2009wt} were originally built by imposing a Galilean symmetry for the scalar field, leaving only five free numerical parameters. We can also cite 
the pure kinetic gravity theory \cite{GUBITOSI2011113}, massive gravity in the non-relativistic limit \cite{deRham:2010ik,deRham:2010kj} and the DBI-Galileon \cite{deRham:2010eu}. 

The DBI-Galileon falls into the class of Brane-world scenarios of extra-dimension theories, where the matter fields are confined on a 4D brane while gravity can propagate into the additional spatial dimensions. Of most interest is the case of a single extra-dimension, as it has been shown that theories with more co-dimensions exhibit ghosts in maximally symmetric spaces 
\cite{Hinterbichler:2010xn}. 
The 5D action includes a volume term for the 4D brane in the 5D bulk which leads to the well-known Dirac-Born-Infeld (DBI) action, 
which can lead to a self-accelerating solution and has been thoroughly studied as a candidate model in the early Universe cosmic inflation paradigm \cite{Silverstein:2003hf,Langlois:2008qf}, for which observational constraints have been derived \cite{Renaux-Petel:2013ppa,Kumar:2015kcb}. Moreover, in the non-relativistic limit the DBI-Galileon model exhibits the Galileon Lagrangians \cite{deRham:2010eu} and gives a physical meaning to their free parameters: the Planck mass in the brane, the Planck mass in the bulk, and so on. 
In particular, the brane tension here plays the role of the cosmological constant which provides a possible interpretation of its nature.

The original probe brane construction from \cite{deRham:2010eu} has been revisited in \cite{Zumalacarregui:2012us} where the matter metric is disformally related to a standard gravitational metric, and in the framework of spontaneous symmetry breaking of the 5D space-time symmetries 
\cite{Goon:2012dy}. The latter publication bridges the gap with Brane-world scenarios developed in the context of quantum field theory and an interpretation of the scalar field as a Nambu-Goldstone boson \cite{Dobado:2000gr,Cembranos:2001my}. The DBI-Galileon model has been studied extensively in the special case of the maximally symmetric bulk geometry \cite{Burrage:2011bt,Goon:2011qf}, and the constraining power of simple cosmological probes in the case of a Minkowski bulk has been explored \cite{Sampurnanand:2012rna} in the context of cosmology. 

In this paper, we develop the DBI-Galileon theory in the non-relativistic limit (Section~\ref{sec:dbigalileon}) and study its dynamics in the Friedmann-Lemaître-Robertson-Walker flat metric as a potential candidate for Dark Energy (Section~\ref{sec:background}). The
stability of cosmological perturbations is explored in Section~\ref{sec:perturbations} and then discussed in Section~\ref{sec:discussion}.

\section{DBI-Galileon in the non-relativistic limit}
\label{sec:dbigalileon}

\subsubsection*{DBI-Galileon \\}

We are interested in the description of a 4-dimensional brane universe embedded in a 5-dimensional bulk from the cosmological perspective. In this context, \cite{deRham:2010eu} shows that the most general action on the brane is given by the 4D Lovelock terms \cite{Lovelock:1971yv} inside the brane and the boundary terms associated to the 5D Lovelock terms: 
\begin{equation}
    \mathcal{S} = \int dx^{4} \sqrt{-g} \left( -\Lambda - M_{5}^{3}K + \frac{M_{P}^{2}}{2}\mathcal{R} - \beta\frac{M_{5}^{3}}{m^{2}}\mathcal{K}_{GB} + \mathcal{L}_{m} \left( \tilde{q}_{\mu\nu}, \psi_{m} \right) \right), \label{eq:dRT action}
\end{equation}
where $g$ is the induced metric on the brane, $\Lambda$ the brane cosmological constant, $K$ the extrinsic curvature of the brane, $\mathcal{R}$ the Ricci scalar on the brane, $\mathcal{K}_{GB}$ the boundary term associated with the bulk Gauss-Bonnet term and $\mathcal{L}_{m}$ the Lagrangian density of matter that lives confined in the brane. The action also introduces the 4D Planck mass $M_{P}$, its 5D counterpart $M_{5}$, their ratio $m=M_{5}^{3}/M_{P}^{2}$, and an arbitrary parameter $\beta$.

Because the action is defined on the 4-dimensional brane, the above quantities are expressed in terms of the induced metric $g$, as opposed to the 5-dimensional bulk metric. We assume that there is a coordinate system $\left(x^{\mu}, y \right)$ in the bulk, Greek letters being defined to span from 0 to 3, such that it is possible to define a normal foliation of the bulk space-time. In this frame, the length element is $ds^{2} = q_{\mu\nu} dx^{\mu}dx^{\nu} + \left( dy \right)^{2}$, with the bulk 4D metric $q$ defined on the tangent space to $y=\mathrm{const}$ hypersurfaces.
The brane position in the bulk is defined by $y = \pi \left( x^{\mu} \right)$ such that we can express the induced metric from the bulk metric $q$:
\begin{equation}
    g_{\mu\nu} = q_{\mu\nu} + \partial_{\mu} \pi \partial_{\nu} \pi \qquad \mathrm{and} \qquad g^{\mu\nu} = q^{\mu\nu} - \gamma^{2} \partial^{\mu} \pi \partial^{\nu} \pi
\end{equation}
with the Lorentz factor $\gamma = \left( 1+ \left( \partial \pi \right)^{2} \right)^{-1/2}$. From this expression, 
we can explicitly write the action \eqref{eq:dRT action} using $q$ and the scalar field $\pi$. In particular, the cosmological constant part of the action leads to a DBI term of the form $-\Lambda \sqrt{-q}\sqrt{1+(\partial \pi)^2}$. In original DBI theories, the first factor is usually the brane tension $f$.
This leads us to interpret $\Lambda$ as related to the brane tension $f$, following $\Lambda = f^{4}$ for unit convenience.

\subsubsection*{Non-relativistic limit \\}

If the derivatives of the field vanish, $\partial_{\mu} \pi = 0$, then we recover standard GR. Because we are interested in the cosmological setting, where predictions from the $\Lambda$CDM model based on GR match precisely a large range of observations, we consider only small corrections to GR. Therefore, we consider the DBI-Galileon in the so-called non-relativistic limit where $\left( \partial \pi \right)^{2} \ll 1$. We can proceed, in addition, to the field redefinition $\pi \rightarrow \varphi= f^{2} \pi $ to have a canonically normalized field.
Up to degree 5 in $\partial \varphi/f^{2}$, we find the following Lagrangian operators for the DBI-Galileon in the non-relativistic limit:
\begin{eqnarray}
    && \mathcal{L}_{f} = 1 + \frac{\mathcal{L}_{2}}{2f^{4}} - \frac{X^{2}}{2} + \ldots, \qquad \qquad \mathcal{L}_{K} = - \frac{\mathcal{L}_{3}}{2f^{6}} - \frac{2X}{f^{6}}\left[ \psi \right] + \ldots, \nonumber \\
    && \mathcal{L}_{\mathcal{R}} = R + \frac{1}{f^{4}} \left( \left[ \Phi \right]^{2} - \left[ \Phi^{2} \right] - \frac{f^{4}X}{2}R - 2R_{\mu\nu}\nabla^{\mu}\varphi \nabla^{\nu}\varphi \right) + \frac{\mathcal{L}_{4}}{4f^{8}} + \ldots, \nonumber \\
    && \mathcal{L}_{\mathcal{K}_{GB}} = \frac{2}{f^{6}} \left( \left( - R_{\mu\nu}\left[ \Phi \right] + 2 R_{\mu\rho} \Phi_{\nu}^{\rho} + R_{\mu\rho\nu\lambda}\Phi^{\rho\lambda} \right) \nabla^{\mu}\varphi \nabla^{\nu}\varphi + \frac{1}{3}\left[ \Phi \right]^{3} - \left[ \Phi \right]\left[ \Phi^{2} \right] + \frac{2}{3}\left[ \Phi^{3} \right] \right. \nonumber \\
    && \left.  - \frac{1}{2}R\left[ \psi \right] \right) + \frac{\mathcal{L}_{5}}{3f^{10}} + \ldots \label{eq:DBI-Galileon Lagrangians}
\end{eqnarray}
We defined in the above expressions the covariant derivative $\nabla$ associated with the metric $q$, the corresponding Riemann tensor $R_{\mu\nu\rho\lambda}$ and its contractions, the tensor $\Phi_{\mu\nu} = \nabla_{\mu}\nabla_{\nu} \varphi$, and the three scalars $\left[ \Phi \right] = \Phi_{\mu}^{\mu}$, $\left[ \psi \right] = \partial^{\mu} \varphi \cdot \Phi_{\mu\nu} \cdot \partial^{\nu} \varphi$ and $X = -\left( \partial \varphi \right)^{2}/2f^{4}$. Furthermore, the $\mathcal{L}_{2 \ldots 5}$ Lagrangian operators are those of the covariant Galileon as defined in \cite{Nicolis:2008in,Deffayet:2009wt} and detailed in \ref{section:Galileon Lagrangians}. Therefore, the theory described here extends the covariant Galileon model which could be of interest as the original version has been proven to be strongly disfavoured by observational data \cite{Leloup:2019fas,Ezquiaga:2017ekz}. We can note that the additional terms in $\mathcal{L}_{\mathcal{R}}$ and $\mathcal{L}_{\mathcal{K}_{GB}}$ compared to the original Galileon Lagrangian operators vanish in the particular case of a flat geometry. 
This link to the covariant Galileon noted, we will stay at leading order in $X$ in the following.

As it was shown in \cite{deRham:2010eu}, the equations of motion are at most of second order, therefore the DBI-Galileon can be expressed as a Horndeski theory \cite{Horndeski:1974wa,Deffayet:2011gz} with the following Horndeski functions $G_i(X)$ that depend on parameters related, in our case, to physical quantities:
\begin{IEEEeqnarray}{rClrCl}
    G_{2} & \equiv & \mathcal{A} \left( \varphi \right) - f^{4} \left( 1 - X + \ldots \right), \qquad \qquad & G_{4} & \equiv & \frac{M_{P}^{2}}{2} \left( 1 - X + \ldots \right), \label{eq:Horndeski G2} \\
    G_{3} & \equiv & \frac{M_{5}^{3}}{f^{2}} \left( X + \ldots \right), & G_{5} & \equiv & -2\beta\frac{M_{5}^{3}}{m^{2}f^{2}} \left( X + \ldots \right). \label{eq:Horndeski G5}
\end{IEEEeqnarray}



\section{Cosmological background evolution}\label{sec:background}

As we are studying the model in the cosmological context, we expand the metric $q$ around a flat FLRW background. The length element $ds^{2} = q_{\mu\nu}dx^{\mu}dx^{\nu}$ reduces at the background level to:
\begin{equation}
    ds^{2} = -dt^{2} + a^{2} \left( t \right)\delta_{ij}dx^{i}dx^{j} \label{eq:FLRW bulk}
\end{equation}

We choose to expand $q$ around a flat FLRW background instead of $g$ because the disformal transformation that link them imply that they differ only by a function of physical time at the background level:

\begin{equation}
    \bar{g}_{\mu\nu} = \bar{q}_{\mu\nu} + \frac{\left( \dot{\bar{\varphi}} \left( t \right) \right)^{2}}{f^{4}} \delta_{\mu 0}\delta_{\nu 0}
\end{equation}
where barred quantities are taken at the background level. Therefore, the geometry on the brane is also of the FLRW type, but with different definitions for the physical time, 
scale factor and expansion history. Because in our case $\dot{\bar{\varphi}} \ll f^{2}$, the expansion history on the 4D slice and on the brane are approximately the same. In addition, equations~\eqref{eq:Horndeski G2}-\eqref{eq:Horndeski G5} determine a self-consistent Horndeski theory of gravity with the metric $q_{\mu\nu}$ and the scalar field $\varphi$ without referring to the induced metric $g_{\mu\nu}$. 
Thus, in the following we apply the well-known techniques used in the context of Horndeski theories.

In the non-relativistic limit, action~\eqref{eq:dRT action} reads:
\begin{equation}\label{eq:action_galdbi}
    \mathcal{S} = \int dx^{4} \sqrt{-q} \left(\mathcal{L}_f + \mathcal{L}_K + \mathcal{L}_{\mathcal{R}} + \mathcal{L}_{\mathcal{K}_{GB}} + \mathcal{L}_{m} \left( q_{\mu\nu}, \psi_{m} \right) \right).
\end{equation}
We define $\Omega_m^0$ and $\Omega_r^0$ the standard present energy density parameters for pressureless matter and radiation respectively, and $\bar{H}$ the normalized Hubble rate $H/H_0$ with $H_0$ the present Hubble constant. Prime symbol denotes the derivative with respect to $\ln a$. We set: 
\begin{equation}
\tilde{x} = \frac{\varphi' H_0}{f^2}, \quad \Omega_\Lambda^0 = \frac{\Lambda}{3 H_0^2 M_P^2}= \frac{f^4}{3 H_0^2 M_P^2}, \quad \eta = \frac{M_5^3}{M_P^2 H_0}, \quad \xi = \frac{\beta}{\eta}, \quad \kappa = \frac{M_P H_0}{f^2}.
\end{equation}
The two Friedmann equations derived from action~\eqref{eq:action_galdbi} are:
\begin{eqnarray}
    && \h^2 = \frac{\Omega_m^0}{a^3} + \frac{\Omega_r^0}{a^4} + \Omega_\Lambda^0 \left( 1 + \frac{1}{2}\h^2 \tilde{x}^2 \right) + \eta \h^4 \tilde{x}^3 - \frac{10}{3} \xi \h^6 \tilde{x}^3 \\[5pt] 
    && \h^2 +\frac{2}{3}\h \h' = - \frac{\Omega_r^0}{3a^{4}} + \Omega_\Lambda^0 \left( 1 - \frac{1}{2}\h^2 \x^2 \right) - \frac{2}{3} \xi\left( 2\h^6 \x^3 + 5\h^5 \x^3 \h'  + 3\h^6 \x^2 \x' \right) \nonumber \\    
    && \qquad\qquad\qquad - \frac{1}{2}\h^4 \x^2 - \h^3 \x^2 \h'  - \frac{2}{3} \h^4 \x \x' + \frac{1}{3} \eta \h^3 \x^2 (\h\x)'
\end{eqnarray}
We see that we recover the $\Lambda$CDM equations when setting $\x$ to zero. The DBI model proposed here is then an extension of the standard model of cosmology,
but with a physical interpretation of the origin of $\Lambda$ as a brane tension, since $\Lambda = f^4$. Using the same methodology and similar notations as in \cite{Stephen-Appleby_2012}, we derive the equation of motion of the scalar field from action~\eqref{eq:action_galdbi}.

For given values of the parameters $\Omega_m^0$, $ \eta$, $\xi$ and $\kappa$, and initial conditions $\tilde x_0$ and $H_0$, the system of equations can be integrated to compute background cosmology observables like the distance moduli of type Ia supernovae. We have used these to put constraints on the parameters and initial conditions, and found sets of parameters that fit observational data at least as well as the concordance model. This is to be expected since the $\Lambda$CDM model is included, by construction.




\section{Stability conditions}\label{sec:perturbations}

To be viable as a description of our Universe, the model has to fulfill stability conditions. 
In the determination of the stability conditions for the DBI-Galileon model in the non-relativistic limit, we use the formalism described in \cite{DeFelice:2011bh} for Horndeski theories, applied to the Horndeski functions \eqref{eq:Horndeski G2} to \eqref{eq:Horndeski G5}.

\subsection*{Tensorial stability conditions}\label{sec:tensorial}

In order to avoid ghosts and Laplacian instabilities, we impose the following constraints on the sign of the kinetic term and on the sign of the gravitational waves speed squared, in the non-relativistic limit and at the lowest order in $\tilde{x}$ :

\begin{equation}
    Q_t \simeq \frac{1}{4} \left( 1 + 2(\h \x)^2 \right) > 0, \qquad \text{and} \qquad c_{t}^{2} \simeq 1 - (\h \x)^2 \geq 0 \label{eq:gw speed}
\end{equation}

In particular, we see that the gravitational wave speed depends on $\x$, and tends to 1 when $\x \ll 1$. Given the very tight constraint on the speed of gravitational waves, equal to the speed of light up to a $\sim 10^{-15}$ difference \cite{LIGOScientific:2017zic,Creminelli:2017sry,Ezquiaga:2017ekz}, this justifies \textit{a posteriori} the relevance of the non-relativistic limit where $\x \ll 1$. Moreover, we see that tensorial perturbations are stable in this limit since $Q_t > 0$.

\subsection*{Scalar stability conditions}

Similar stability conditions apply to the scalar degrees of freedom, here including the scalar perturbations of matter components \cite{DeFelice:2011bh}.
At the lowest order in $\x$, we get:
\begin{equation}
Q_s \simeq \frac{3}{2} (\Omega_\Lambda^0 - \h^2) \x^2 > 0 \qquad \text{and} \qquad c_{s}^{2} \simeq 1 + \frac{2 \left( \eta \h^{2}\x' - \h\h' - 2\xi\h^{4}\x' \right)}{3 \left( \Omega_{\Lambda}^{0} - \h^{2} \right)} \geq 0 \label{eq:Qs DBI-galileon}
\end{equation}

With $\x \ll 1$ and ignoring instabilities of the scalar perturbations, a fit of the DBI-Galileon model at the background level to distance data leads to cosmological parameters close to the standard model ones: $\Omega_m^0 \approx 0.3$ and $\Omega_\Lambda^0 \approx 0.7$ \cite{Planck2018}. Therefore, from the first Friedmann equation, we get $\Omega_\Lambda^0 < \h^2$ for all sets of parameters in agreement with cosmological distance observations. As $Q_s \leq 0$, the DBI-Galileon model contains scalar instabilities unless it reduces to GR, in which case $Q_{s} = 0$ because the scalar field vanishes. One way to avoid this would be to add a spatial curvature to the metric with an energy density of at least $\Omega_{k}^{0} \sim 0.3$. However, we anticipate that this would lead to a bad fit of cosmological observations from the proximity of the model to $\Lambda$CDM.

\section{Discussion}\label{sec:discussion}

\subsection*{Physical interpretation}

From definition \eqref{eq:Qs DBI-galileon}, we see that the dominant terms come from $\G2$
($\Omega_\Lambda^0$ term) and $\G4$
($\h^2$ term). The competition between these, i.e. between the DBI and the Ricci scalar terms,
leads to the ghost-like behaviour in cosmology.
The DBI action will have the effect of stretching the brane towards an extremal surface, whereas the Ricci scalar term on the brane will tend to make the brane contract on itself from the effect of space-time curvature. However, in the non-relativistic limit,
the Ricci scalar term destabilizes the scalar field perturbations and the stretching effect from $\Lambda$
is not strong enough to counterbalance it, leading to an instantaneous decay of the vacuum state.

Because the DBI-Galileon action is the most general one for a 4D probe brane in a 5D bulk, we expect this statement to be quite general for all such models
studied under our assumptions, i.e. in the non-relativistic limit and in a standard cosmological setting.

In the context of late-time cosmology where matter is present, there might be direct coupling to the scalar field whose interactions with matter could help stabilize the scalar perturbations. However, we checked that the DBI-Galileon action in the non-relativistic limit is invariant under a disformal transformation of the metric, following the treatment of \cite{Bekenstein:1992pj} and \cite{Stephen-Appleby_2012}, at the lowest order in $X$. Therefore, direct couplings to matter can always be absorbed by a redefinition of the DBI-Galileon parameters, and are therefore already included in the previous study that exhibit ghosts.

In conclusion, the only way to evade this ghostly behaviour in cosmology would be 
to include the full relativistic dynamics of the theory ($\x \sim 1$). However, higher order terms would need to be included in \eqref{eq:gw speed} and
observable deviations of the speed of gravitational waves $c_t$ from $c$ should occur. On the other hand, we can view the full DBI-Galileon as an effective theory valid only at cosmological scales, where the speed of gravitational waves has not been probed \cite{deRham:2018red,Ezquiaga:2018btd}. Indeed, the constraint on $c_{t}$
from the observation of GW170817 in coincidence with GRB170817A \cite{LIGOScientific:2017zic} is only valid on the small scales probed by LIGO and Virgo. A modification of the dispersion relation of gravitational waves at small scales from operators present in the UV complete theory could allow $c_{t} \neq c$ on cosmological scales while being compatible with current astrophysical observations. Waiting for the next generation of gravitational wave interferometers, in particular LISA, which will be able to probe this relation at larger scales \cite{Barausse:2020rsu}, this possibility remains open.

\subsection*{Generalization}

The DBI-Galileon is an
example of a
shift-symmetric Horndeski theory, a subclass
of Horndeski theories
invariant under a shift symmetry of the scalar field $\varphi \rightarrow \varphi + c$ \cite{Sotiriou:2013qea,Sotiriou:2014pfa}. In these, the
Horndeski functions are restricted to depend only on $X$.
To make the non-relativistic limit apparent, we expand these
functions around GR:
\begin{IEEEeqnarray}{rClrCl}
    G_{2} & \equiv & \Lambda + \sum_{n=1}^{+\infty} g_{2}^{\left( n \right)} X^{n}, \qquad \qquad & G_{4} & \equiv & \frac{M_{P}^{2}}{2} + \sum_{n=1}^{+\infty} g_{4}^{\left( n \right)} X^{n} \\
    G_{3} & \equiv & \sum_{n=1}^{+\infty} g_{3}^{\left( n \right)} X^{n}, & G_{5} & \equiv & \sum_{n=1}^{+\infty} g_{5}^{\left( n \right)} X^{n}
\end{IEEEeqnarray}

The constant terms in $G_{3}$ and $G_{5}$ do not appear in the expansion as they lead to total derivative terms.

From these, we can compute the stability conditions at the lowest order:
\begin{IEEEeqnarray}{rClrCl}
    Q_s & \simeq & \frac{X}{H^{2}} \left( \g2^{\left( 1 \right)} + 6H^{2} \g4^{\left( 1 \right)} \right) > 0, \qquad \qquad \qquad \qquad \qquad & Q_t & \simeq & \frac{M_{P}^{2}}{4} > 0 \\
    c_{s}^{2} & \simeq & 1 + \frac{2\Ddot{\phi} \g3^{\left( 1 \right)} + 4\Dot{H}\g4^{\left( 1 \right)} + 2\Ddot{\phi}H^{2}\g5^{\left( 1 \right)}}{\g2^{\left( 1 \right)} + 6H^{2} \g4^{\left( 1 \right)}} \geq 0, & c_t^{2} & \simeq & 1 \geq 0
\end{IEEEeqnarray}

The two tensorial conditions are automatically satisfied in this context. On the other hand, the stability conditions for scalar perturbations give a simple inequality involving the parameters of the Taylor expansion, that can be easily checked at the background level. In particular, this gives a very simple formulation of the scalar no-ghost condition, independent of $X$, for shift-symmetric Horndeski theories in the non-relativistic limit:
\begin{equation}
     \g2^{\left( 1 \right)} + 6H^{2} \g4^{\left( 1 \right)} > 0
\end{equation}

In the context of the brane galileon, where $\g4^{\left( 1 \right)} = -M_{P}^{2}/2$ and $\g2^{\left( 1 \right)} = \Lambda$, this is equivalent to the inequality that we found in \eqref{eq:Qs DBI-galileon}, which is never fulfilled in flat space:
\begin{equation}
    \Lambda - 3M_{P}^{2}H^{2} > 0 \quad \Leftrightarrow \quad \Omega_{\Lambda}^{0} > \h^{2}
\end{equation}

\section{Conclusion}\label{sec:conclusion}

We described the DBI-Galileon theory of a four-dimensional brane evolving in a 5D bulk space-time in the non-relativistic limit where its local kinetic energy is small compared to its tension. In this framework, the DBI-Galileon corresponds to an expansion around standard GR, and therefore around standard $\Lambda$CDM in the cosmological context. As such, it is naturally compatible with cosmological data in the non-relativistic limit, provided the effect of the scalar field is small enough.
In contrast to $\Lambda$CDM, however, the free parameters of the DBI-Galileon model acquire a physical meaning from its construction. In particular, the interpretation of the cosmological constant is linked to the brane tension. We derived the equations driving the evolution of the late-time Universe around a spatially flat FLRW cosmological background and studied the stability of scalar and tensorial perturbations. We found that the model exhibits a fatal ghostly behaviour for scalar perturbations  around the FLRW background. From there, we derived the corresponding stability conditions for shift-symmetric Horndeski theories in the non-relativistic limit in the cosmological context and found very simple formulations for these conditions, generalizing our results obtained for the DBI-Galileon model.

\section*{Acknowledgements}

We would like to thank Marc Besançon, Arnaud de Mattia and Vanina Ruhlmann-Kleider for their comments on the present paper. We also want to thank David Langlois for useful and interesting comments and suggestions.

\section*{References}
\bibliographystyle{iopart-num}
\bibliography{bibliography}

\appendix
\section{Galileon Lagrangians}
\label{section:Galileon Lagrangians}

Following the notations introduced in Section~\ref{sec:dbigalileon}, the Covariant Galileon Lagrangians, defined in \cite{Deffayet:2009wt}, that enter the Lagrangians of the DBI-Galileon theory are:
\begin{eqnarray}
    \mathcal{L}_{2} & = & -2X \\
    \mathcal{L}_{3} & = & -2X \left[ \Phi \right] \\
    \mathcal{L}_{4} & = & -2X \left( 2 \left[ \Phi \right]^{2} - 2 \left[ \Phi^{2} \right] + RX \right) \\
    \mathcal{L}_{5} & = & -2X \left( \left[ \Phi \right]^{3} - 3 \left[ \Phi \right] \left[ \Phi^{2} \right] + 2 \left[ \Phi^{3} \right] - 6 \, G_{\nu\rho} \nabla_{\mu} \phi \nabla^{\mu\nu} \phi \nabla^{\rho} \phi \right)
\end{eqnarray}

\end{document}